\documentclass[amsmath, amssymb, aps, twocolumn]{revtex4-1}
\usepackage{graphicx}
\usepackage{amsmath}
\usepackage{bm}
\usepackage{mathrsfs}
\usepackage{booktabs}

\begin{document}

\title{Large spin-orbit splitting and weakly-anisotropic superconductivity revealed with single-crystalline noncentrosymmetric CaIrSi$_3$}

\author{G. Eguchi$^1$}
\email{geguchi@scphys.kyoto-u.ac.jp}
\author{H. Wadati$^2$}
\author{T. Sugiyama$^3$}  \author{E. Ikenaga$^4$} 
\author{S. Yonezawa$^1$} \author {Y. Maeno$^1$}
\affiliation{$^1$Department of Physics, Graduate School of Science, Kyoto University, Kyoto 606-8502, Japan \\
$^2$Department of Applied Physics and Quantum-Phase Electronics Center (QPEC), University of Tokyo, Hongo, Tokyo 113-8656, Japan \\
$^3$Research Center for Synchrotron Light Applications, Kyushu University, Kasuga 816-8580, Japan \\
$^4$Japan Synchrotron Radiation Research Institute (JASRI)/SPring-8, Kouto, Mikazuki 679-5198, Japan}

\date{\today}

\begin{abstract}
We report normal and superconducting properties of the Rashba-type noncentrosymmetric compound CaIrSi$_3$, using single crystalline samples with nearly 100\% superconducting volume fraction. The electronic density of states revealed by the hard x-ray photoemission spectroscopy can be well explained by the relativistic first-principle band calculation. This indicates that strong spin-orbit interaction indeed affects the electronic states of this compound. The obtained $H-T$ phase diagram exhibits only approximately 10\% anisotropy, indicating that the superconducting properties are almost three dimensional. Nevertheless, strongly anisotropic vortex pinning is observed.

\begin{description}
\item[PACS numbers]
74.25.-q, 81.10.Dn, 79.60.-i, 71.20.Be
\end{description}

\end{abstract}


\maketitle

\section{Introduction}
Superconductivity in the absence of inversion symmetry has been attracting much attention for its potential exotic superconducting phenomena~\cite{Gorkov2001PRL,Agterberg2007PRB}. In bulk materials, such superconductivity is realized in crystals without inversion centers, namely in the so-called noncentrosymmetric superconductors (NCSCs)~\cite{Bauer2004PRL}. The lack of the inversion symmetry results in two important features. Firstly, the superconducting state cannot be classified either as a pure spin-singlet or a pure spin-triplet pairing any more, but is  a singlet-triplet mixed pairing. Then, the superconducting gap function is generally expressed as $\hat{\varDelta}_{\bm{k}}=\{\varDelta_1\psi(\bm{k})+\varDelta_2\bm{d}(\bm{k})\cdot{\hat{\bm{\sigma}}}\}i\hat{\sigma}_y$. Here, $\psi(\bm{k})$ and $\bm{d}(\bm{k})$ are the normalized scalar superconducting gap function for the spin-singlet component and the normalized vector gap function for the spin-triplet component, respectively; $\varDelta_1$ and $\varDelta_2$ are the gap magnitudes, including $\varDelta_2=0$ (spin-singlet state) or $\varDelta_1=0$ (spin-triplet state) as special cases; and $\hat{\sigma}$ is the Pauli matrices.

Secondly, the lack of inversion symmetry results in electronic energy-band splitting due to the antisymmetric spin-orbit interaction (ASOI). The ASOI term in the Hamiltonian is given by the inner product of the dimensionless $g$-vector ${\bm{g}}({\bm{k}})$ and the $\bm{k}$-dependent electron spin ${\bm{\sigma}}({\bm{k}})$: $H_{\rm{ASOI}}=\alpha\Sigma_k{\bm{g}}({\bm{k}})\cdot {\bm{\sigma}}({\bm{k}})$. Here $\alpha$ characterizes the strength of the ASOI. The ASOI leads to an energy shift of $\pm \alpha|{\bm{g}}({\bm{k}})|$, which can be interpreted as an effective $\bm{k}$-dependent Zeeman splitting of the electron spins~\cite{Kaur2005PRL}. It has been revealed that ${\bm{d}}({\bm{k}}) \parallel {\bm{g}}({\bm{k}})$ is energetically favored for $\alpha \gg |\varDelta_{\bm{k}}|$. Furthermore, $\psi(\bm{k})$ and $\bm{d}(\bm{k})$ belong to the same irreducible representation of the crystalline point group, and ${\bm{d}}({\bm{k}})$ is expressed as ${\bm{d}}({\bm{k}})=\psi(\bm{k}){\bm{g}}({\bm{k}})$~\cite{Sigrist2006MMM,Hayashi2006PRB}. This indicates that the superconducting energy gap $\varDelta_{\bm{k} \pm}$ takes the form $\varDelta_{\bm{k} \pm}=\psi(\bm{k})\{\Delta_1 \pm \Delta_2| \bm{g}(\bm{k})|\}$ on each split energy band. 
 
One of the typical and well-studied ASOIs is of the Rashba-type, whose $g$-vector has the form ${\bm{g}}({\bm{k}}) \propto (k_y, -k_x, 0)$~\cite{Agterberg2007PRB}. Because of this simple form, NCSCs having the Rashba-type ASOI are favorable for the investigation of the novel superconducting state. In fact, a number of unusual superconducting phenomena originating from the parity mixing or the ASOI have been proposed: e.g., helical vortex state and novel magneto electric effect~\cite{Kaur2005PRL,Fujimoto2007JPSJ051008}. For observation of any of such phenomena, single crystalline sample is crucially important, because the novel effects are likely to be canceled out in polycrystals. 

Naively speaking, the Rashba-type ASOI is expected to be realized in crystals in which inversion symmetry is broken only along one direction. 
To our knowledge, only seven Rashba-type NCSCs with single crystals are reported up until today: $R$Pt$_3$Si ($R$=La, Ce)~\cite{Yasuda2004JPSJ,Takeuchi2007JPSJ}, $R$RhSi$_3$ and $R$IrSi$_3$($R$=La, Ce)~\cite{Kimura2005PRL,Okuda2007JPSJ,Settai2008JPSJ}, and CeCoGe$_3$~\cite{Settai2007IJMPB}. The four Ce-based NCSCs indeed exhibit exotic superconducting behavior. However, they also exhibit antiferromagnetic ordering in the vicinity of the superconducting phase. Thus, it is difficult to distinguish effects originating from the ASOI from effects resulting from active $f$-electrons. In contrast, the other three La-based NCSCs exhibit conventional metallic behavior in their normal states and weak-coupling full-gap behavior in their superconducting states. This fact may suggest importance of the active $f$-electrons for realization of the novel phenomena. However, a few NCSCs having different crystal structures indeed exhibit novel superconducting phenomena, even though active $f$-electrons are not present~\cite{Yuan2006PRL,Bauer2010PRB_Mo3Al2C}. Thus, further investigations of non-magnetic, $f$-electron-free Rashba-type NCSCs are important.

Here, we report our success in growing single crystals of the Ca-based NCSC CaIrSi$_3$, which has the same crystal structure as $R$IrSi$_3$ ($R$=La, Ce). CaIrSi$_3$ exhibits superconductivity below the critical temperature $T_{\rm{c}}=3.6$~K~\cite{Oikawa2008JPSmeeting}, which is the highest $T_{\rm{c}}$ among the known Rashba-type NCSCs. We previously reported that CaIrSi$_3$ is a non-magnetic, fully-gapped superconductor based on studies with polycrystalline samples~\cite{Eguchi2011PRB,Eguchi2012JPSJ}. Single crystalline samples allow examinations of various predicted exotic phenomena as well as investigation of basic information of superconductivity, such as the gap anisotropy and the mixing ratio of the singlet and triplet components. In this paper, we describe the synthesis of single crystalline CaIrSi$_3$, and its normal and superconducting properties. We also present the electronic density of states (DOS) near the Fermi energy $E_{\rm{F}}$ revealed by bulk sensitive hard x-ray photoemission spectroscopy (HAXPES), and by relativistic first-principle calculations~\cite{Schwarz2003}. By comparing the results of HAXPES and the band calculations, we revealed existence of strong SOI in this material. The compound can be a model material having strong SOI with 5$d$ orbital characters.

\section{Single crystal growth}

\begin{figure}
\centering
\includegraphics[width=\columnwidth, clip]{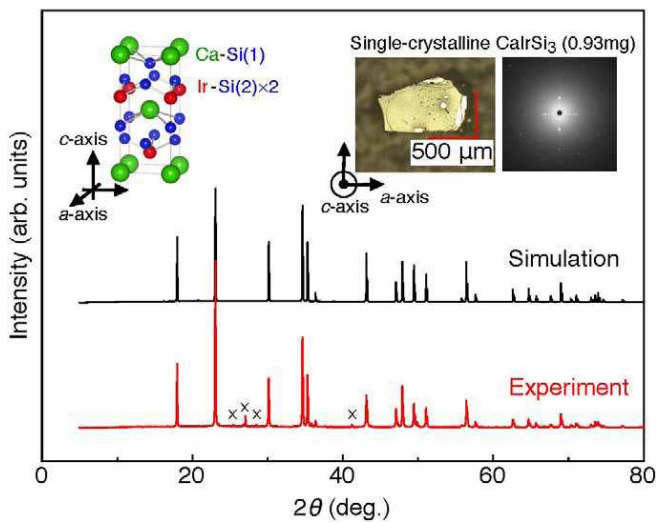}
\caption{(Color online) Experimental and simulated powder x-ray diffraction spectra of CaIrSi$_3$. The small impurity peaks indicated by $\times$ marks are attributed to a by-product CaIr$_3$Si$_7$~\cite{Eguchi2011PRB}, whose grains are distinct from CaIrSi$_3$ crystals. The upper-left inset illustrates the crystal structure of CaIrSi$_3$. This schematic was produced by the software VESTA~\cite{Momma2011AC}. A optical microscope picture of a single crystal and its backscattered Laue photograph are also shown in the upper-right inset.}
\label{crystal}
\end{figure}

Single crystalline CaIrSi$_3$ is synthesized by a combination of the arc melting~\cite{Eguchi2011PRB} and self-flux methods. Firstly, polycrystalline CaIrSi$_3$ was obtained by the arc melting of a pellet made of a powder mixture of CaSi (99.9\%), Ir (99.99\%), and Si (99.999\%) with the molar ratio Ca:Ir:Si~=~3:1:4.7. Secondly, small single crystals of CaIrSi$_3$, which were typically less than 0.1~mm in sizes, were isolated from the arc-melted sample by dissolving a by-product CaSi$_x$ into 3\% hydrochloric acid. Thirdly, these small crystals were mixed with a sufficient amount, typically twenty times of the Ir-molar mass of the crystals, of the CaSi, Ir, and Si powders with the molar ratio Ca:Ir:Si~=~3:1:4.7 in Ar atmosphere, then pressed into a pellet. The small single crystals should play a role of seed crystals. The pellet was put into an alumina crucible, which was then capsulated into a stainless steel container in an Ar atmosphere~\cite{Kihou2010JPSJ}. The container was heated up to 1340$^\circ\mathrm{C}$, cooled to 800$^\circ\mathrm{C}$ by 2$^\circ\mathrm{C}$/h, then quenched. Many pieces of single crystalline CaIrSi$_3$ grown up to the size 0.2-1.4~mm are obtained along with a slight amount of by-product CaIr$_3$Si$_7$.

A powder x-ray diffraction spectrum (Bruker AXS D8 ADVANCE) of crushed single crystals is presented in Fig.~\ref{crystal}. The obtained spectrum well agrees with the simulation. An optical microscope picture of a crystal with the typical size of $0.5 \times 0.5 \times 0.5$~mm$^3$ is presented in the upper-right inset. A backscattered Laue photograph of the presented crystal is also shown. This photograph was taken with an apparatus~(RIGAKU RASCO-BLII) in which a charge-coupled-device~(CCD) camera takes the image of a Laue pattern projected onto a fluorescent screen. The clear four-fold symmetry in the Laue picture indicates that the surface shown corresponds to the basal $ab$~plane.

\section{Hard x-ray photoemission spectrum}

\begin{figure}
\centering
\includegraphics[width=\columnwidth, clip]{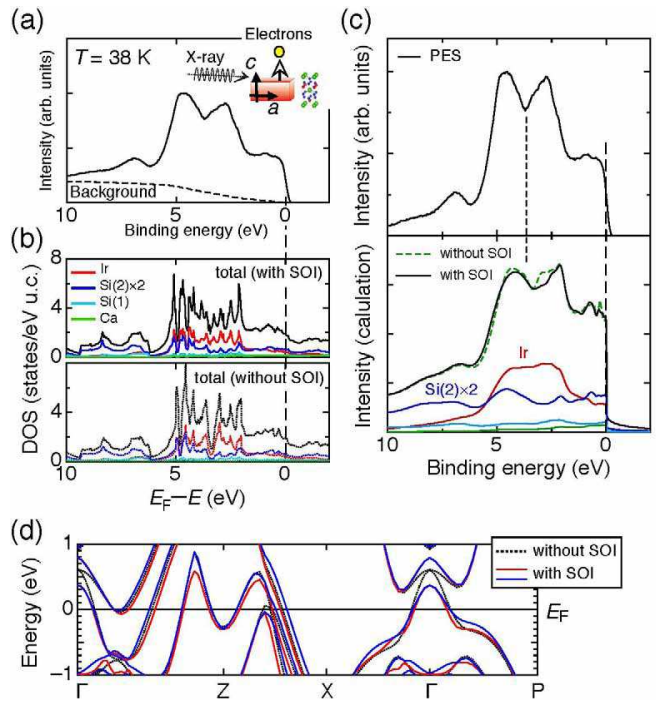}
\caption{(Color online) (a) Hard x-ray photoemission spectrum of CaIrSi$_3$ near $E_{\rm{F}}$. A schematic of the experimental setup is presented in the inset. (b) Calculated density of states (DOS) per unit cell by the local density approximation (LDA) band calculation with and without spin-orbit interaction (SOI). Partial DOS for each atom is also presented. (c) Comparison between the background-subtracted experimental spectrum and the calculated spectrum with an assumed life-time broadening. The calculation with SOI well describes the experimental result than that without SOI does. Contribution of each atom to the spectrum is also presented based on the calculation with SOI. (d) Band dispersion with/without SOI. Spin-orbit splitting of the two-fold spin degeneracy near $E_{\rm{F}}$ is $0.05\sim0.3$~eV.}
\label{PES}
\end{figure}

A photoemission spectrum around $E_{\rm{F}}$ revealed by the bulk sensitive HAXPES is presented in Fig.~\ref{PES}(a). The measurement was performed at the beamline BL47XU at SPring-8 (Japan)~\cite{Kobayashi2009NIMA}. The setup of the measurement is depicted in the inset: incident photons with an energy of 7.9399~keV have their polarization plane perpendicular to the sample surface. The DOS deduced from the full potential linearize augmented plane wave (FLAPW) calculation performed by WIEN2k package with/without SOI are presented in Fig.~\ref{PES}(b) together with the partial DOS for each atom. Our calculation well reproduces the previous calculations in Refs.~\cite{Bannikov2010,Kaczkowski2011JAC}. The calculated DOS at $E_{\rm{F}}$ per unit cell~(u.c.) is 1.94~states/eV\hspace{1pt}u.c., and the Sommerfeld coefficient $\gamma_0$ evaluated from the DOS is 4.6~mJ/mol-u.c.\hspace{1pt}K$^2$. This value is consistent with the previous reports~\cite{Bannikov2010,Kaczkowski2011JAC}. Note that $\gamma_0$ per formula unit~(f.u.), which is compared with the experimental results later, is 2.3~mJ/mol-f.u.\hspace{1pt}K$^2$, because one unit cell contains two CaIrSi$_3$ formula unit. From the DOS, we simulate the HAXPES spectrum by assuming a Lorentzian-type lifetime broadening with the energy-dependent line width (FWHM=$0.2|E-E_{\rm{F}}|$)~\cite{Wadati2005PRB}, and compared it in Fig.~\ref{PES}(c) with the background-subtracted experimental spectrum. The calculation with the SOI explains the experimental result better than that without the SOI: In particular, the shoulder-like structure at approximately 3~eV for the latter is absent in the experimental spectrum. 
The fact indicates that strong SOI indeed affects the electronic state of CaIrSi$_3$. 

The calculated band dispersions with/without the SOI are shown in Fig.~\ref{PES}(d). The four-fold degeneracy at the symmetric $\Gamma$ point is split into two by $\sim0.42$~eV due to the SOI. Note that this energy split at the $\Gamma$ point ($\bm{k}=0$) is caused by the symmetric SOI, which can be finite regardless of the crystal symmetry. The value of the $\Gamma$-point splitting is consistent with a previous study~\cite{Kaczkowski2011JAC}. The value is also comparable to that of CeIrSi$_3$: 0.4~eV~\cite{Jeong2010SSC}, in which Ce 4$f$ orbitals dominantly contribute to the electronic conduction. The remaining two-fold spin degeneracy is split by $\sim0.05$-$0.3$~eV at less symmetric $k$ points due to the ASOI.  The band splitting results in approximately 10\% DOS difference ($\delta N \sim 0.1$) between the ASOI-split Fermi surfaces. As presented in Figs.~\ref{PES}(b)(c), the Ir-5$d$ orbital contributes by 20-50\% to the total DOS near $E_{\rm{F}}$. Thus, the influence of the strong SOI discussed above is attributed to the Ir orbitals. 

\section{Normal-state and Superconducting properties}

\begin{figure}
\centering
\includegraphics[width=8cm, clip]{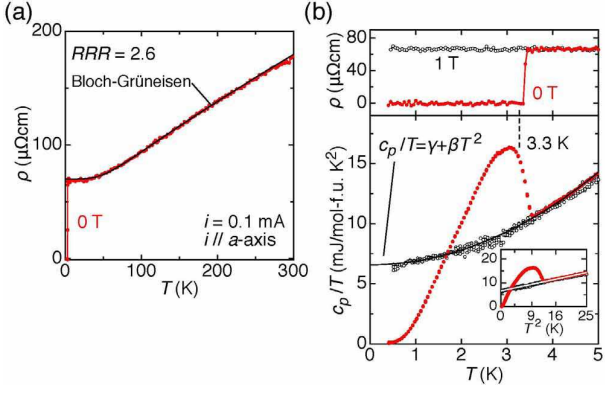}
\caption{(Color online) (a) Temperature dependence of the resistivity $\rho$ of the single crystalline CaIrSi$_3$. The solid line is a fit by the conventional Bloch-Gr{\"u}neisen model. (b) Low-temperature resistivity and specific heat divided by temperature, $c_p /T$, at 0~T and 1~T ($H \parallel a$). The broken vertical line indicates $T_{\rm{c}}=3.3$~K based on the entropy conservation, and the solid line is a fit by the conventional Debye-Sommerfeld model: $c_p/T=\gamma +\beta T^2$. The inset is a $c_p/T$-vs-$T^2$ plot, indicating that the model well describes the normal-state behavior.}
\label{cp_etc}
\end{figure}

The temperature dependence of the resistivity $\rho$ is presented in Fig.~\ref{cp_etc}(a), while the low temperature $\rho$ and specific heat $c_p$ of the same sample are presented in Fig.~\ref{cp_etc}(b). The measurements were performed with a commercial apparatus~(Quantum Design, PPMS) down to 0.35~K with a $^3$He refrigerator. For the specific heat measurements, three crystals from the same batch were added to achieve sufficient experimental resolution (total: 7.301~mg). Their crystal axes were determined individually by the Laue photography. The onset temperature of the specific-heat jump in zero field is 3.55~K, which is consistent with the results for polycrystalline samples~\cite{Eguchi2011PRB}. The thermodynamic $T_{\rm{c}}$ determined by the entropy conservation criteria, is 3.3~K. The superconductivity is suppressed by a field of 1~T ($H \parallel a$). Both the specific heat and resistivity in the normal state are magnetic-field independent within the experimental resolutions. The residual resistivity $\rho_0$ of the present sample is 68~$\mu\Omega$cm. The residual resistivity ratio $RRR \equiv \rho(300~{\rm{K}})/\rho(5~{\rm{K}})$ is 2.6. Both $\rho_0$ and $RRR$ are comparable to that of polycrystalline samples~\cite{Eguchi2011PRB}. The temperature dependence of $\rho$ is well fitted by the conventional Bloch-Gr{\"u}neisen formula with the transport Debye temperature 307~K. This fact indicates dominance of electron-phonon scatterings. The transport Debye temperature is also consistent with that of polycrystalline samples~\cite{Eguchi2012JPSJ}.

The normal state $c_p$ below 5~K is well described by the conventional Debye-Sommerfeld model: $c_p=\gamma T + \beta T^3$, where $\gamma$ is the electronic specific heat~(Sommerfeld) coefficient and $\beta$ is the phononic specific heat coefficient. We obtain $\gamma=6.6$ mJ/mol-f.u.\hspace{1pt}K$^2$ and $\beta=0.31$~mJ/mol-f.u.\hspace{1pt}K$^4$ from the fitting to the normal state data between 0.35 and 5~K. The Debye temperature $\varTheta_{\rm{D}}$ calculated from $\beta=(12/5)\pi^4N_{\rm{A}}N_{\rm{f.u.}}k_{\rm{B}}/\varTheta_{\rm{D}}^3$ is 314~K. Here, $N_{\rm{A}}$ is the Avogadro number, $N_{\rm{f.u.}}=5$ is the number of atoms per formula unit, and $k_{\rm{B}}$ is the Boltzmann constant. The $\varTheta_{\rm{D}}$ value is consistent with that obtained from the resistivity data. We obtain the electron-phonon coupling constant $\lambda_{\rm{el-ph}}$ as 0.56 from McMillan's formula $T_{\rm{c}}=(\varTheta_{\rm{D}}/1.45){\rm{exp}}[-1.04(1+\lambda_{\rm{el-ph}})/\{\lambda_{\rm{el-ph}}-\mu^*(1+0.62\lambda_{\rm{el-ph}})\}]$~\cite{McMillan1967PR} with $\varTheta_{\rm{D}}=314$~K and the Coulomb pseudo potential $\mu^*=0.13$. The obtained small value of $\lambda_{\rm{el-ph}}$ indicates that a weak-coupling superconductivity is realized in this compound, being consistent with the electronic specific heat results described below. By comparing the experimental $\gamma$ and the calculated $\gamma_0$ values, the electronic mass enhancement $\gamma/\gamma_0$ is estimated to be 2.9. Assuming that this mass enhancement originates from the electron-phonon interaction and electron-electron interaction, i.e., $\gamma/\gamma_0=(1+\lambda_{\rm{el-ph}})(1+\lambda_{\rm{el-el}})$, we obtain the mass enhancement factor due to the electron-electron correlation $\lambda_{\rm{el-el}}=0.84$.

\begin{figure}
\centering
\includegraphics[width=8cm, clip]{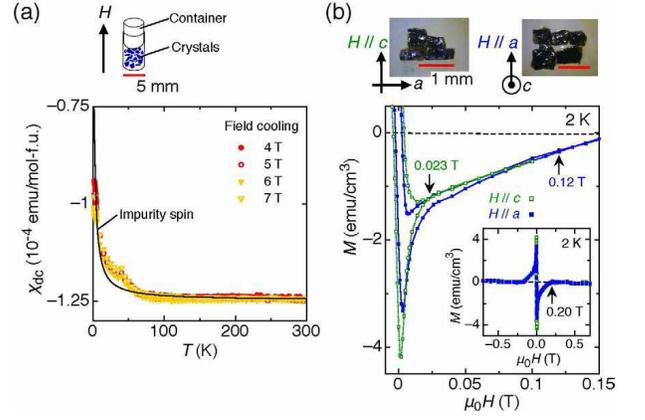}
\caption{(Color online) (a) Temperature dependence of the field-cooled dc susceptibility $\chi_{\rm{dc}}$ in several magnetic fields. The increases of $\chi_{\rm{DC}}$ below 80~K, and the small peaks around 50~K are attributed to impurity spin, and oxygen molecules, respectively (see the text). The measurement setup is also depicted. (b) Magnetization $M$ vs. external field $H$ curves at 2~K for $H \parallel a$ axis (filled symbols) and $H \parallel c$ axis (open symbols). The irreversibility field for each field direction is indicated by arrows. The inset is the full $M-H$ curves. The dotted line with a slight negative slope represents the normal state diamagnetism at 2~K. The onset of the magnetic shielding is 0.2~T at 2~K for both field directions. The measurement setups of the samples are also presented.}
\label{dcchi}
\end{figure}

The temperature dependence of the field-cooled dc susceptibility $\chi_{\rm{dc}}=M/H$ down to 2~K in several magnetic fields are presented in Fig.~\ref{dcchi}(a).  Here, $M$ is the magnetization measured with a commercial magnetometer~(Quantum Design, MPMS-XL), and $H$ is the external magnetic field. The measurements were performed with many single crystals without alignment of the crystalline direction. The setup is depicted at the top of Fig.~\ref{dcchi}(a). An additional peak around 50~K is attributed to oxygen molecules that remains in the sample space. The temperature dependence of $\chi_{\rm{dc}}$ is explained by the sum of a constant term $\chi_0$ and the temperature-dependent spin paramagnetism. In order to obtain $\chi_0$ as well as to estimate the upper limit of the number of paramagnetic spins $N$, we fitted the data with the formula $\chi_{\rm{dc}}=\chi_0+(N\mu_{\rm{B}}/H){\rm{tanh}}(\mu_{\rm{B}}H/k_{\rm{B}}T)$, by assuming the paramagnetic spin to be 1/2. Here $\mu_{\rm{B}}$ is the Bohr magnetron. We obtain $\chi_0=-1.25\times10^{-4}$~emu/mol-f.u. and $N=3.3\times 10^{-5}$~mol/mol-f.u.. The small $N$ value indicates that the observed spin paramagnetism is not intrinsic, i.e. attributable to impurities. The constant term $\chi_0$ is attributed to the bulk $\chi_{\rm{DC}}$ of CaIrSi$_3$, and the negative $\chi_0$ value indicates that the compound is a diamagnet. Meanwhile, the Pauli paramagnetic susceptibility of CaIrSi$_3$ estimated from the FLAPW calculations is $\chi_{\rm{P}}=0.31\times10^{-4}$~emu/mol-f.u. The observed diamagnetism is attributable to the dominance of the core diamagnetism. 

The superconducting $M-H$ curves at 2~K for dc magnetic fields $H \parallel c$ and $H \parallel a$, obtained from separate measurements with a few aligned single crystals, are presented in Fig.~\ref{dcchi}(b). Photos of the experimental setups are presented at the top of Fig.~\ref{dcchi}(b); we arranged the same crystals so that the difference of demagnetization factors between the measurements is small. Indeed, the difference in their superconducting magnetic shielding in the Meissner state is less than 30\%. The inset is the whole $M-H$ loops, indicating that the magnetic shielding occurs below $\pm$0.20~T for both field directions. The loops, exhibiting typical behavior of type-II superconductivity with weak vortex pinning, provide evidence that CaIrSi$_3$ is a type-II superconductor. The type-II superconductivity is also indicated by our previous study with polycrystalline samples~\cite{Eguchi2011PRB}. As presented in the figure, the irreversible field is 0.12~T for $H\parallel a$, and 0.023~T for $H\parallel c$. This anisotropic behavior indicates that vortex pinning for $H\parallel a$ is stronger than that for $H\parallel c$. Such anisotropic pinning was reproducibly observed in other measurements with a single piece of a crystal from the same batch and with several pieces of crystals from another batch. We note that no anisotropy in the normal state susceptibility was observed within our experimental resolution. 

\begin{figure}
\centering
\includegraphics[width=\columnwidth, clip]{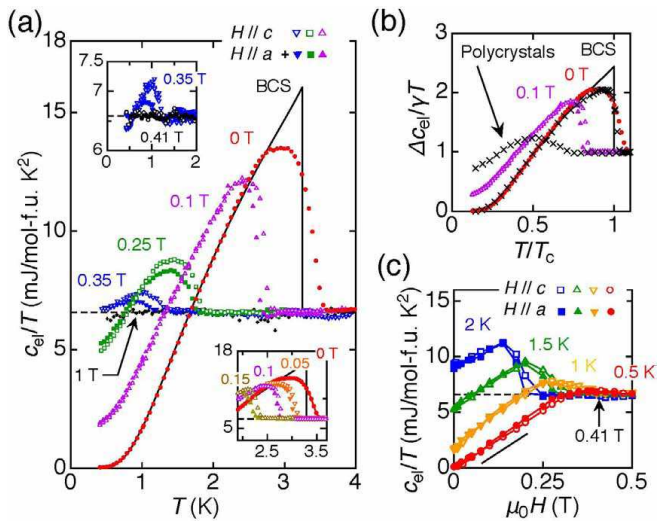}
\caption{(Color online) (a) Temperature dependence of the electronic specific heat divided by temperature $c_{\rm{el}}/T$ for $H \parallel a$ (closed symbols) and for $H \parallel c$ (open symbols). We use the relation $c_{\rm{el}}/T=c_p/T-\beta T^2$ to deduce the electronic contribution. The detailed $c_{\rm{el}}/T$ around 0.1~T and 0.4~T are presented in the insets. The horizontal broken line presents the obtained Sommerfeld coefficient $\gamma=6.6$ mJ/mol-f.u.\hspace{1pt}K$^2$. The solid curve is a BCS curve obtained from $\gamma$ and the thermodynamic $T_{\rm{c}}$. (b) The comparison of the normalized $\varDelta c_{\rm{el}}/\gamma T$ vs. $T/T_{\rm{c}}$ for the single crystals and the polycrystals~\cite{Eguchi2011PRB}. For the polycrystalline data, $\gamma$ has been corrected for the residual $c_{\rm{el}}/T$ at 0~T (such $\gamma$ is denoted as $\gamma_{\rm{s}}$ in Fig.~5 of \cite{Eguchi2011PRB}). (c) Magnetic field dependence of $c_{\rm{el}}/T$ extracted from temperature-sweep data.}
\label{cel}
\end{figure}

The temperature dependence of the electronic specific heat divided by temperature $c_{\rm{el}}/T=c_p/T-\beta T^2$ in several fields are presented in Fig.~\ref{cel}(a). The measurements for $H \parallel a$ and $H \parallel c$ were performed using the same crystals. At 0~T, the residual term of $c_{\rm{el}}/T$ for $T\rightarrow 0$ is almost absent, indicating that the superconducting volume fraction is nearly 100\% for these samples. The thermodynamic critical field $\mu_0H_{\rm{c}}(0)$ evaluated from the relation $\mu_0H_{\rm{c}}^2(0)/2=-\gamma T^2_{\rm{c}}/2+\int_0^{T_{\rm{c}}}c_{\rm{el}}(T) dT$ is 0.028~T. The theoretical $c_{\rm{el}}/T$ based on the weak-coupling BCS model~\cite{Muhlschlegel1959ZP} presented in the figure well reproduces the observed temperature dependence below 2.7~K. The agreement indicates that the superconducting gap is finite on the entire Fermi surface. Note that similar full-gap behavior of the specific heat has been reported for isostructural polycrystalline BaPtSi$_3$~\cite{Bauer2009PRB} and LaRhSi$_3$~\cite{PhysRevB.83.064522}. The deviation of the $c_{\rm{el}}/T$ from the BCS curve above 2.7~K seems to be attributed to a $T_{\rm{c}}$ distribution within the samples. However, we will point out two tendencies opposite to the ordinary behavior. As presented in the lower inset of Fig.~\ref{cel}(a), the transition in 0~T is broader than that in 0.1~T. This is observed with another measurement with single crystalline samples from different batch (not shown). In addition, the specific-heat jump for the single crystals in 0~T is broader than that for polycrystals, as shown in the comparison of normalized $\Delta c_{\rm{el}}/\gamma T$ between the single crystal and the polycrystals in the Fig.~\ref{cel}(b). These features will be discussed again in the next section. Note that the sharpening of the specific-heat jump in 0.1~T is not observed in polycrystals, probably due to the presence of the grain boundaries in polycrystals. As presented in the upper inset of Fig.~\ref{cel}(a), The specific heat jump is visible in 0.35~T but not in 0.41~T for both field directions.

The magnetic field dependence of $c_{\rm{el}}/T$ extracted from temperature-sweep data are presented in Fig.~\ref{cel}(c). The linear $H$ dependence of $c_{\rm{el}}/T$ at 0.5~K, guided by the solid line, indicates nearly isotropic superconducting gap~\cite{PhysRevB.70.100503}.

\begin{figure}
\centering
\includegraphics[width=\columnwidth, clip]{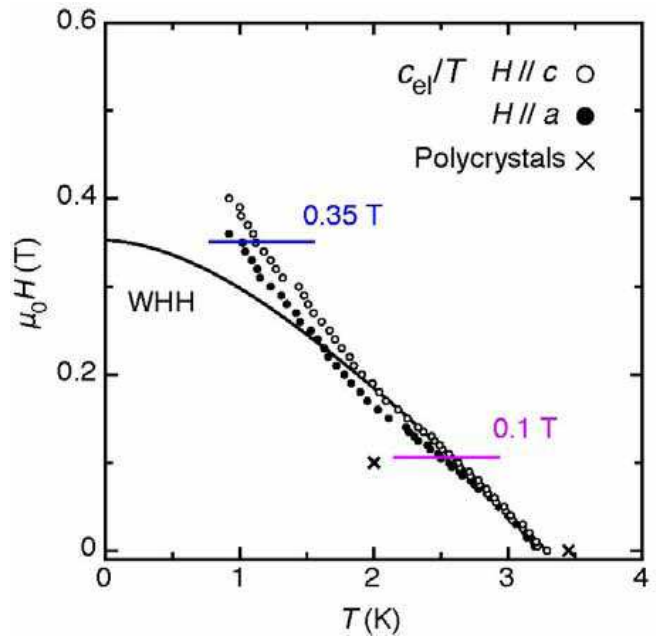}
\caption{(Color online) Superconducting $H$-$T$ phase diagram based on the thermodynamic $T_{\rm{c}}$ for $H\parallel a$ (filled symbols) and $H\parallel c$ (open symbols). The solid curve represents the conventional Werthamer-Helfand-Hohenberg (WHH) curve determined from the $H\parallel a$ data. The thermodynamic $T_{\rm{c}}$ values of the polycrystalline sample~\cite{Eguchi2011PRB} are also presented.}
\label{HT}
\end{figure}

The $H-T$ phase diagram deduced from the thermodynamic $T_{\rm{c}}$ for each field direction is presented in Fig.~\ref{HT}. For comparison, the thermodynamic $T_{\rm{c}}$ of polycrystals evaluated from the data reported in Ref.~\cite{Eguchi2011PRB} are also presented. The conventional Werthamer-Helfand-Hohenberg (WHH) curve for the $H_{\rm{c2}}(T)$ in the dirty limit, giving $\mu_0H_{\rm{c2}}^{\rm{WHH}}(0)=-0.693T_{\rm{c}}[{\rm{d}}(\mu_0H_{\rm{c2}})/{\rm{d}}T]|_{T=T_{\rm{c}}}=0.35$~T, is also presented~\cite{Helfand1966PRpart3}. The WHH curve well agrees with the experimental result for $H\parallel c$ above 2~K. However, the observed $H_{\rm{c2}}$ becomes substantially larger than $\mu_0H_{\rm{c2}}^{\rm{WHH}}(T)$ at lower temperatures for both field directions. In addition, $H_{\rm{c2}}$ for the $c$ direction is higher than that for the $a$ direction. This is also evident in the specific-heat data in Fig.~\ref{cel}(a). The ratio $H_{\rm{c2}}^{\parallel c}$/$H_{\rm{c2}}^{\parallel a}(T)$ is approximately 1.1, which is almost invariant below 3~K. Such small anisotropy indicates a nearly three-dimensional superconducting nature.

The Ginzburg-Landau~(GL) parameter $\kappa_{\rm{GL}}$ estimated from $\mu_0H_{\rm{c2}}^{\parallel a}=0.375$~T at 0.9~K is 10.2, using the relation $\kappa=H_{\rm{c2}}^{\parallel a}/(\sqrt{2}H_{\rm{c}})$ and $\mu_0H_{\rm{c}}=0.026$~T at 0.9~K. The value of $H_{\rm{c}}$ is evaluated from the $c_{\rm{el}}/T$ data in 0~T (see e.g.~\cite{Tinkham2nd}). The obtained $\kappa_{\rm{GL}}$ value indicates type-II superconductivity, which is consistent with the results of $M$-$H$ curves in Fig.~\ref{dcchi}(b). 
The lower critical field $\mu_0H_{\rm{c1}}^{\parallel a}=0.0018$~T at 0.9~K is estimated from the relation $H_{\rm{c}}^2=H_{\rm{c1}}^{\parallel a}H_{\rm{c2}}^{\parallel a}$. The GL superconducting coherence lengths $\xi_{a}$ and $\xi_{c}$ are estimated to be $\xi_c=31$~nm and $\xi_a=29$~nm at 0.9~K from the relation $\mu_0H_{\rm{c2}}^{\parallel a}=\Phi_0/2\pi\xi_{a}\xi_{c}$ and $\mu_0H_{\rm{c2}}^{\parallel c}=\Phi_0/2\pi\xi_{a}^2$. Note that the $\mu_0H_{\rm{c2}}^{\parallel c}$ at 0.9~K is 0.40~T. The corresponding penetration depth is readily obtained from $\kappa_{\rm{GL}}=\lambda/\xi$. All of these values are consistent with those of polycrystals~\cite{Eguchi2011PRB}.

\section{Discussion}

According to the group theory, the superconducting gap symmetry of CaIrSi$_3$ belongs to the $C_{\rm{4v}}$ point group, which is the same as that of CePt$_3$Si~\cite{Sigrist2006MMM}. There are five possible superconducting pairing symmetries allowed for this point group. The observed fully-gapped behavior indicates that the most symmetric $A_1$ state with $\Delta_1>\Delta_2$ is realized, because this is the only case having a fully-gapped superconducting energy gap function. The gap function of the $A_1$ state with the Rashba-type ASOI is expressed as $\Delta_{\pm}(\theta_{\bm{k}})=\Delta_1\pm \Delta_2 {\rm{sin}}\theta_{\bm{k}}$ for the pair of ASOI-split bands. Here $\theta_{\bm{k}}$ is the angle between the $c$ axis and $\bm{k}$. Note that the superconductivity becomes multi-gapped if $\Delta_2$ is finite. Although the low-temperature $c_{\rm{el}}/T$ behavior in 0~T is consistent with the simplest BCS curve, a possibility of a weak band dependence of the superconducting gap is not excluded.

As mentioned above, the superconducting transition at 0~T is broader than that at 0.1~T. Furthermore, the observed $H-T$ curves deviate from the conventional WHH curve below 2~K. These feature may well be explained as a result of a distribution in both $T_{\rm{c}}$ and $H_{\rm{c2}}$. In fact, the observed behavior is also reproducible by assuming a two-$T_{\rm{c}}$ BCS model, having one domain with lower $T_{\rm{c}}$ and higher $H_{\rm{c2}}$, and  another domain with higher $T_{\rm{c}}$ and lower $H_{\rm{c2}}$. On the other hand, the observed behavior can also be explained as a result of multi-gapped superconductivity with a weak band dependence of the gap. In the present case, there are two possible multi-gapped superconductivity: one originating from multiple Fermi surfaces, or one from a finite $\Delta_2$. The deviation from the simplest BCS approximation is worth investigation, and a high-resolution photoemission spectroscopy or a scanning tunneling spectroscopy would give crucial indications. 

Another interesting feature in the observed superconducting behavior is the anisotropic pinning observed in the $M-H$ curves: the irreversible field at 2~K for $H\parallel a$ is approximately five times larger than that for $H\parallel c$. Such anisotropic behavior cannot be explained by a simple pinning model with random defects and impurities. Thus, the anisotropic vortex pinning probably indicate an anisotropic distribution of lattice imperfections. CaIrSi$_3$ can have twin boundaries of crystalline domains with opposite directions of the Rashba field, as discussed in CePt$_3$Si~\cite{Mukuda2009JPSJ}. Such twin boundaries would run both perpendicular to an in-plane axis and perpendicular to the $c$-axis. The anisotropic pinning may be caused by these twin boundaries. Note that novel vortex behavior occurring at the twin boundary, i.e., a fractional vortex, is discussed for CePt$_3$Si for $H \parallel a$, in order to explain the observed extremely slow vortex dynamics~\cite{Iniotakis2008JPSJ}. Such interesting vortex phenomena might be related to the observed anisotropic pinning in CaIrSi$_3$. 

It has been theoretically predicted that the helical vortex state, which is unique to NCSCs, can be stable in the presence of finite $\delta N$~\cite{Agterberg2007PRB,Kaur2005PRL}. The helical vortex state has a superconducting gap function $\varDelta(\bm{r})=|\varDelta| e^{i\bm{q}_{\rm{h}}\cdot\bm{r}}$ with $\bm{q}_{\rm{h}} \propto \hat{c} \times \bm{H}$. This is analogous to the Fulde-Ferrell (FF) state~\cite{Fulde1964PR}, which exhibits a spatial phase modulation of $\varDelta(\bm{r})$. Note that the finite $\bm{q}_{\rm{h}}$ originates from the asymmetric distortion of a Fermi surface induced by $H$, while the finite $\bm{q}_{\rm{FF}}$ of the FF state originates from the electron pairing between the Zeeman-split energy bands. For another difference, the helical vortex state can be realized down to near $H=0$ whereas the FF state can be stable only in high fields near the Pauli-limitting field $H_{\rm{P}}$. For the helical vortex state, an enhancement of low-temperature $H_{\rm{c2}}$ associated with the emergence of a spatially modulated superconducting gap amplitude, or modulated vortex states, are discussed near $H_{\rm{P}}$~\cite{Agterberg2007PRB,Matsunaga2008PRB}. However, this is not relevant to the case of CaIrSi$_3$ because $\mu_0H_{\rm{P}}(0)\sim 6$~T is sufficiently larger than the actual $H_{\rm{c2}}$. Here, $\mu_0H_{\rm{P}}(0)$ is estimated from the relation $\mu_0H_{\rm{P}}(0)/T_{\rm{c}}=1.84$~T/K. 
Nevertheless, other signatures related to the helical vortex state may be observed in CaIrSi$_3$. For example, the observed $H_{\rm{c2}}$ enhancement in low temperatures is possibly related to properties of the helical vortex state at $H \ll H_{\rm{P}}$. As well as the sample improvement, further investigation especially in the lower temperature region is necessary.

\begin{table}[h]
\caption{Physical properties of CaIrSi$_3$ obtained by this study. Here $\gamma$ is the electronic specific heat coefficient per formula unit, $\varTheta_{\rm{D}}$ is the Debye temperature, $\chi_0$ is the bulk dc susceptibility per formula unit, $T_{\rm{c}}$ is the thermodynamic superconducting transition temperature, $\mu_0H_{\rm{c}}(0)$ is the thermodynamic critical field, $\lambda_{\rm{el-ph}}$ is the electron-phonon coupling constant, $\lambda_{\rm{el-el}}$ is the electronic mass enhancement due to the electron-electron correlation, $\mu_0H_{\rm{c2}}^{\parallel a}$ is the upper critical field at 0.9~K, $\kappa_{\rm{GL}}$ is the Ginzburg-Landau parameter, $E_{\rm{F}}$ is the Fermi energy, $N(E_{\rm{F}})$ is the density of states per unit cell, and $\chi_{\rm{P}}$ is the Pauli paramagnetic susceptibility per formula unit. The latter three values  are obtained from the relativistic first principle calculations.}
	\begin{ruledtabular}
	\begin{tabular}{cc}
		$\gamma$ & 6.6~mJ/mol-f.u.\hspace{1pt}K$^2$ \\
		 $\varTheta_{\rm{D}}$ & 314~K \\
		 $\chi_0$ &  -1.25$\times 10^{-4}$~emu/mol-f.u. \\
		 $T_{\rm{c}}$ & 3.3~K \\
		 $\mu_0H_{\rm{c}}(0)$ & 0.028~T \\
		 $\lambda_{\rm{el-ph}}$ & 0.56  \\
		 $\lambda_{\rm{el-el}}$ & 0.84  \\
		 $\mu_0H_{\rm{c2}}^{\parallel a} $(at 0.9~K) & 0.375~T \\
		 $\kappa_{\rm{GL}}$ (at 0.9~K) & 10.2 \\ \midrule
		 $E_{\rm{F}}$ & 8.26~eV \\
		 $N(E_{\rm{F}})$ & 1.94~states/eV\hspace{1pt}u.c. \\
		 $\chi_{\rm{P}}$ & 0.31 $\times 10^{-4}$~emu/mol-f.u. \\
	\end{tabular}
	\end{ruledtabular}
\label{t1}
\end{table}

\section{Summary}
We revealed the existence of strong SOI on CaIrSi$_3$ by HAXPES, and the nearly three-dimensional, fully gapped superconducting nature with anisotropic vortex pinning, using single crystalline samples. The physical properties evaluated in this study is summarized in the Table~\ref{t1}. We emphasize that CaIrSi$_3$ is one of the simplest Rashba-type NCSC having strong SOI without active $f$-electrons. It is expected that the single crystalline samples would provide unique oppotunities for investigations of novel superconducting phenomena related to the lack of the inversion symmetry.

\begin{acknowledgements}
We thank Taichi Matsuda, Hiroaki Ikeda, Y. Yanase, S. Fujimoto, D.C. Peets, M. Kriener, F. Kneidinger, and E. Bauer for fruitful discussions. This work is supported by a Grant-in-Aid for the Global COE program ``The Next Generation of Physics, Spun from Universality and Emergence'' from the Ministry of Education, Culture, Sports, Science, and Technology (MEXT) of Japan, and by the ``Topological Quantum Phenomena'' Grant-in Aid for Scientific Research on innovative Areas from MEXT of Japan. It is also supported by the Japan Society for the Promotion of Science (JSPS) through the ``Funding Program for World Leading Innovative R\&D on Science and Technology (FIRST Program)", initiated by the Council for Science and Technology Policy (CSTP). The synchrotron radiation experiments at SPring-8 were performed under the approvals of the Japan Synchrotron Radiation Research Institute (2011B1710 and 2012A1624). G.E. is also supported by JSPS.
\end{acknowledgements}

\bibliographystyle{apsrev4-1_nocomma}
\bibliography{Preload,myrefs}

\end{document}